\documentclass[journal]{IEEEtran}
\usepackage[T1]{fontenc}
\usepackage{textcomp}
\usepackage[utf8]{inputenc}
\usepackage{amsmath}
\usepackage{graphicx}
\usepackage[bookmarks=true,bookmarksnumbered=true,bookmarksopen=true,bookmarksopenlevel=1,
 breaklinks=false,pdfborder={0 0 0},pdfborderstyle={},backref=false,colorlinks=false]
 {hyperref}
\hypersetup{pdftitle={Your Title},
 pdfauthor={Your Name},
 pdfpagelayout=OneColumn, pdfnewwindow=true, pdfstartview=XYZ, plainpages=false}

\makeatletter
\usepackage[caption=false,font=footnotesize]{subfig}
\usepackage{tabularx}

\makeatother

\begin{document}
\title{Channel Knowledge Map Construction: Recent Advances and Open Challenges}
\author{Zixiang Ren, Juncong Zhou, Jie Xu, Ling Qiu, Yong Zeng, Han Hu, Juyong
Zhang, and Rui Zhang \thanks{Z. Ren, J. Zhou, and J. Xu are with the School of Science and Engineering
(SSE), the Shenzhen Future Network of Intelligence Institute (FNii-Shenzhen),
and the Guangdong Provincial Key Laboratory of Future Networks of
Intelligence, The Chinese University of Hong Kong (Shenzhen), Longgang,
Shenzhen, Guangdong 518172, China (e-mail: rzx66@mail.ustc.edu.cn;
juncongzhou@link.cuhk.edu.cn; xujie@cuhk.edu.cn). J. Xu is the corresponding
author.}\thanks{L. Qiu is with the Key Laboratory of Wireless-Optical Communications,Chinese
Academy of Sciences, School of Information Science and Technology,
University of Science and Technology of China, Hefei 230027,China
(e-mail: lqiu@ustc.edu.cn).}\thanks{Y. Zeng is with the National Mobile Communications Research Laboratory,
School of Information Science and Engineering, Southeast University,
Nanjing 210096, China, and also with the Purple Mountain Laboratories,
Nanjing 211111, China (e-mail: yong\_zeng@seu.edu.cn).}\thanks{H. Hu is with the School of Information and Electronics, Beijing Institute
of Technology, Beijing 100081, China (e-mail: hhu@bit.edu.cn).}\thanks{J. Zhang is with the School of Mathematical Sciences, University of
Science and Technology of China, Hefei 230027, China. (e-mail: juyong@ustc.edu.cn).}\thanks{R. Zhang is with the Department of Electrical and Computer Engineering,
National University of Singapore, Singapore 117583 (e-mail:elezhang@nus.edu.sg).}}
\maketitle
\begin{abstract}
Channel knowledge map (CKM) has emerged as a pivotal technology for
environment-aware wireless communications and sensing, which provides
\textit{a priori} location-specific channel knowledge to facilitate
network optimization. Efficient CKM construction is an important technical
problem for its effective implementation. This article provides a
comprehensive overview of recent advances in CKM construction. First,
we examine classical interpolation-based CKM construction methods,
highlighting their limitations in practical deployments. Next, we
explore image processing and generative artificial intelligence (AI)
techniques, which leverage feature extraction to construct CKMs based
on environmental knowledge. Furthermore, we present emerging wireless
radiance field (WRF) frameworks that exploit neural radiance fields
or Gaussian splatting to construct high-fidelity CKMs from sparse
measurement data. Finally, we outline various future research directions
in real-time and cross-domain CKM construction, as well as cost-efficient
deployment of CKMs. 
\end{abstract}


\IEEEpeerreviewmaketitle{}

\section{Introduction}

Channel knowledge map (CKM) has emerged as a cornerstone technology
for sixth-generation (6G) wireless networks, enabling a paradigm shift
from environment-ignorant network design to proactive and intelligent
environment-aware optimization \cite{zeng2021toward}. In general,
CKM is a representation framework that provides the functional mapping
from the spatial positions of transceivers to environment-specific
channel knowledge such as channel gains, channel state information
(CSI), reference signal strength indicator (RSSI), and multi-path
delays/Dopplers/angles-of-arrival (AoAs)/angles-of-departure (AoDs)
\cite{howmuchdata,zeng2024tutorial}. Depending on application scenarios,
CKMs can be represented in various forms, like spatial databases,
images, and even neural network models. By providing location- and
environment-specific channel knowledge {\it a priori}, CKM has emerged
as an effective tool to facilitate channel prediction, enable predictive
wireless communications, and support environment-aware beam tracking
and network deployment optimization \cite{wu2023environment,10962296}.

CKMs can be classified into base station (BS)-to-any (B2X) and any-to-any
(X2X) types according to the input dimension of transceiver positions.
B2X CKM provides channel knowledge associated with a fixed-position
BS to support BS-centric communications, in which the input may correspond
to the two-dimensional (2D) user position in, e.g., terrestrial networks
with all users at the same altitude or three-dimensional (3D) positions
of users for, e.g., low-altitude networks with users distributed in
3D space. By contrast, X2X CKM takes both transmitter (Tx) and receiver
(Rx) positions as input, allowing the construction of channel knowledge
between arbitrary Tx-Rx pairs. Depending on whether 2D or 3D Tx/Rx
positions are considered, the X2X CKM can be constructed in 4D or
6D formats, respectively. Among them, 6D X2X CKM offers the most comprehensive
channel knowledge, which can thus be employed for complex tasks such
as the predictive resource allocation for device-to-device (D2D) communications,
unmanned aerial vehicle (UAV) trajectory planning, and environment-aware
network deployment.

The construction and utilization of CKMs are two stages for the realization
of environment-aware wireless communication and sensing, and the successful
construction of accurate CKM is of foremost significance \cite{10962296}.
CKM construction refers to the process of generating a spatially resolved
representation of channel knowledge by leveraging prior measurement
data and/or physical environmental information. For example, ray tracing
is a commonly adopted deterministic channel modeling technique that
can be used for CKM construction based on physical environment information.
Ray tracing simulates electromagnetic wave propagation by tracking
multiple paths between the Tx and Rx, accounting for reflection, diffraction,
and scattering based on the environment’s geometry. Although theoretically
precise, ray tracing suffers from prohibitive computational complexity
and a strong dependence on accurate environmental information, such
as geometry, dielectric properties, and material characteristics,
thus limiting its practical implementation for large-scale and real-time
applications. Recently, with the emergence of cellular networks and
the Internet of Things, the massive number of connected devices can
provide location-specific channel measurements. Combined with physical
environment information, these measurement data can enable the efficient
construction of CKMs via exploiting emerging data-driven approaches,
achieving a balance between construction accuracy and computational
efficiency.

In the literature, representative CKM construction methods include
interpolation, image processing and generative artificial intelligence
(AI), as well as wireless radiance field (WRF) approaches. First,
interpolation is widely used for constructing B2X CKMs, which estimates
channel knowledge at unmeasured positions by using sparse measurement
data from a limited number of reference points via techniques such
as kernel regression, matrix completion, or Kriging \cite{sun2022propagation,li2022channel}.
These methods exploit spatial channel correlations and rely on the
stationarity of the environment, thus making them unsuitable for high-dimensional
X2X CKMs in complex environments. Next, image processing is another
widely adopted approach for 2D B2X CKM construction based on sparse
or low-resolution channel measurements, which treats the CKM as a
gray-scale image, enabling the use of image processing and deep learning
techniques like convolutional neural networks (CNNs) and super-resolution
reconstruction \cite{I2IInpainting}. Furthermore, generative AI techniques
such as diffusion models have demonstrated strong capability for high-fidelity
construction of 2D B2X CKM, which exploit the denoising mechanism
to enable resolution enhancement and map completion (e.g., \cite{wu2025ckmimagenet,fu2025ckmdiff}). 

More recently, inspired by the developments in computer graphics,
wireless radiance field (WRF) frameworks have emerged as a new CKM
construction method, by leveraging neural radiance fields (NeRF) and
3D Gaussian splatting (3DGS) to capture spatial and angular variations
of wireless channels in continuous 3D space \cite{zhao2023nerf2,wen2024wrf}.
A WRF provides a continuous function mapping from 3D Tx/Rx position
and propagation direction to channel knowledge, capturing environment-induced
effects like reflection and scattering via proper models like neural
networks or 3D Gaussian ellipsoids. On the one hand, NeRF-based WRF
learns implicit volumetric representations by using multi-layer perceptrons
(MLPs) to fit geometry and direction-dependent channel features. In
contrast, 3DGS offers explicit parameterization through Gaussian ellipsoids,
enabling efficient, high-resolution modeling of spatial variations
and directional dependencies for complex 3D communication environments.
These methods achieve high-fidelity CKM construction in 3D B2X scenarios
by efficiently learning the joint spatial–angular channel mapping
\cite{zhao2023nerf2,wen2024wrf}, and can also be exploited for constructing
the most comprehensive 6D X2X CKMs via the recently proposed bidirectional
wireless Gaussian splatting (BiWGS) \cite{zhou}. 

In light of the growing trend, this article provides a systematic
overview on the recent advancements in CKM construction by particularly
focusing on the three closely related approaches based on interpolation,
image processing and generative AI, as well as the more recent WRF,
respectively. For each approach, we first review representative methods
and their underlying principles, and then discuss their application
scenarios with different case studies. 

\section{Interpolation for CKM Construction }

Interpolation-based approaches have long been a conventional technique
for constructing CKMs \cite{zeng2024tutorial}, providing an efficient
means of estimating channel knowledge at unmeasured locations from
sparse measurements. In this case, CKMs are typically organized as
geo-tagged databases, where channel knowledge parameters are indexed
by geographic coordinates. The process of building such databases
generally begins with data collection across the target area by using
sparsely deployed sensors or devices. Interpolation is then performed
to fill the measurement voids from sparsely distributed measurement.

Within this framework, Kriging, kernel regression, and matrix completion
are three widely used interpolation methods for CKM construction \cite{sun2022propagation}.
Kriging models spatial stationarity via variograms and produces unbiased
estimators that account for both sample distance and spatial configuration.
Kernel regression applies weighted kernels, e.g., Gaussian functions,
to highlight the influence of nearby samples. Matrix completion exploits
the low-rank structure of grid-sampled channel data, recovering missing
entries by minimizing a cost function under low-rank constraints.

\begin{figure}[tbh]
\centering\includegraphics[scale=0.25]{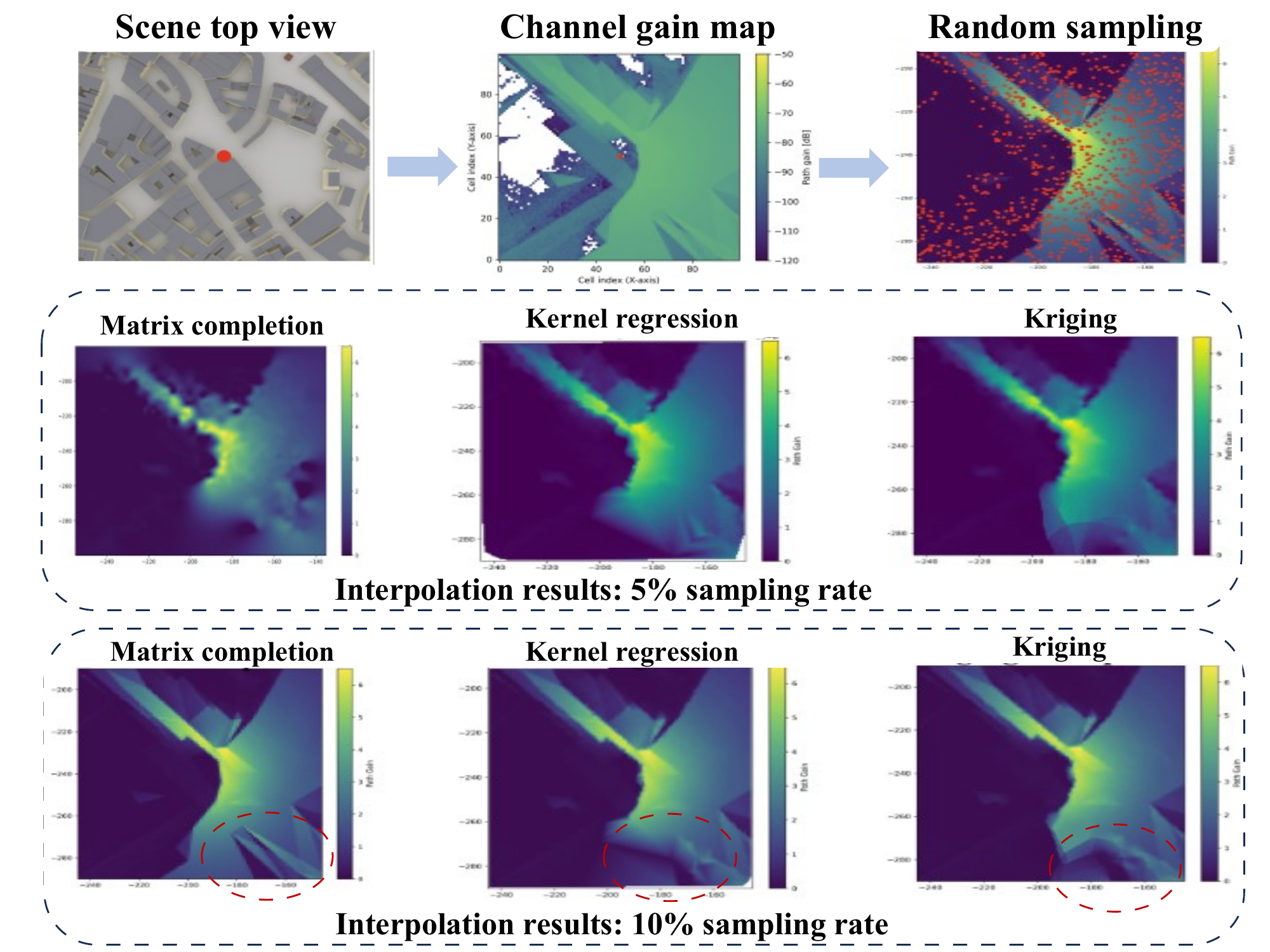}\caption{\label{fig:Interpolation-based-path-gain}Interpolation-based channel
gain map construction.}
\end{figure}

Fig. \ref{fig:Interpolation-based-path-gain} shows a case study for
the construction of channel gain map on a geo-realistic Munich cityscape,
in which the ray tracing data generated via Sionna are considered
as ground truth\footnote{Sionna is an open-source GPU-accelerated library that supports end-to-end
wireless system design and CKM simulation, which is available at https://developer.nvidia.cn/sionna. }. The $100\textrm{ m}\times100\textrm{ m}$ region is discretized
into 10,000 grids (1 m resolution), where exhaustive ray tracing via
Sionna's radio map solver requires $>10{{}^8}$ computations per grid.
We evaluate the construction performance by interpolation-based CKM
construction methods via matrix completion, kernel regression, and
Kriging, under sparse sampling rates of 5\% and 10\%, respectively.
It is observed that CKMs constructed with a 5\% sampling rate exhibit
noticeably lower detail clarity and lower overall fidelity than those
constructed with a 10\% sampling rate. This indicates that lower sampling
rates significantly degrade construction quality, with all methods
exhibiting noticeable artifacts and loss of fine-scale details. It
is also observed that distinct diffraction edges are abrupt in the
construction results, with particularly prominent features located
within the red-circled region. These interpolation approaches depend
heavily on spatial stationarity assumptions, which limit their ability
to accurately capture localized irregularities such as sharp diffraction
edges. Among them, matrix completion demonstrates relatively better
preservation of structural patterns due to its low-rank representation.
Kernel regression and Kriging yield smoother constructions but overly
blur diffraction boundaries due to averaging. 

Despite its wide application, interpolation-based CKM construction
suffers from multiple limitations that hinder its successful implementation
in complex wireless environments. First, although interpolation-based
methods exhibit satisfactory performance in 2D B2X CKM construction,
extending this approach to 3D B2X and 4D/6D X2X scenarios presents
substantial challenges, particularly in modeling vertical channel
variations and capturing multi-dimensional spatial characteristics.
Second, the interpolation-based CKM construction methods rely on the
assumption of spatial stationarity of CKMs, making them highly vulnerable
to non-stationary propagation conditions such as blockage, diffraction,
and complex urban layouts, which significantly degrade the CKM construction
accuracy. Third, interpolation-based methods require densely sampled
and large-scale channel measurements to ensure accurate construction,
and insufficient sampling density may significantly degrade the construction
quality. 

\section{Image Processing and Generative AI for CKM Construction}

Recently, AI techniques have found abundant applications in image
processing and generation, enabling powerful capabilities in feature
extraction, denoising, and super-resolution. This motivates the representation
of CKMs as 2D spatial images, in which each pixel corresponds to channel
knowledge such as amplitude, phase, or path loss at that position.
Such a representation allows the application of advanced computer
vision and generative AI techniques to enhance CKM construction, capture
fine-grained spatial variations, and synthesize high-fidelity maps.

\subsection{Deep Learning-based Architecture for CKM Construction}

\begin{figure}[tbh]
\centering

\includegraphics[scale=0.2]{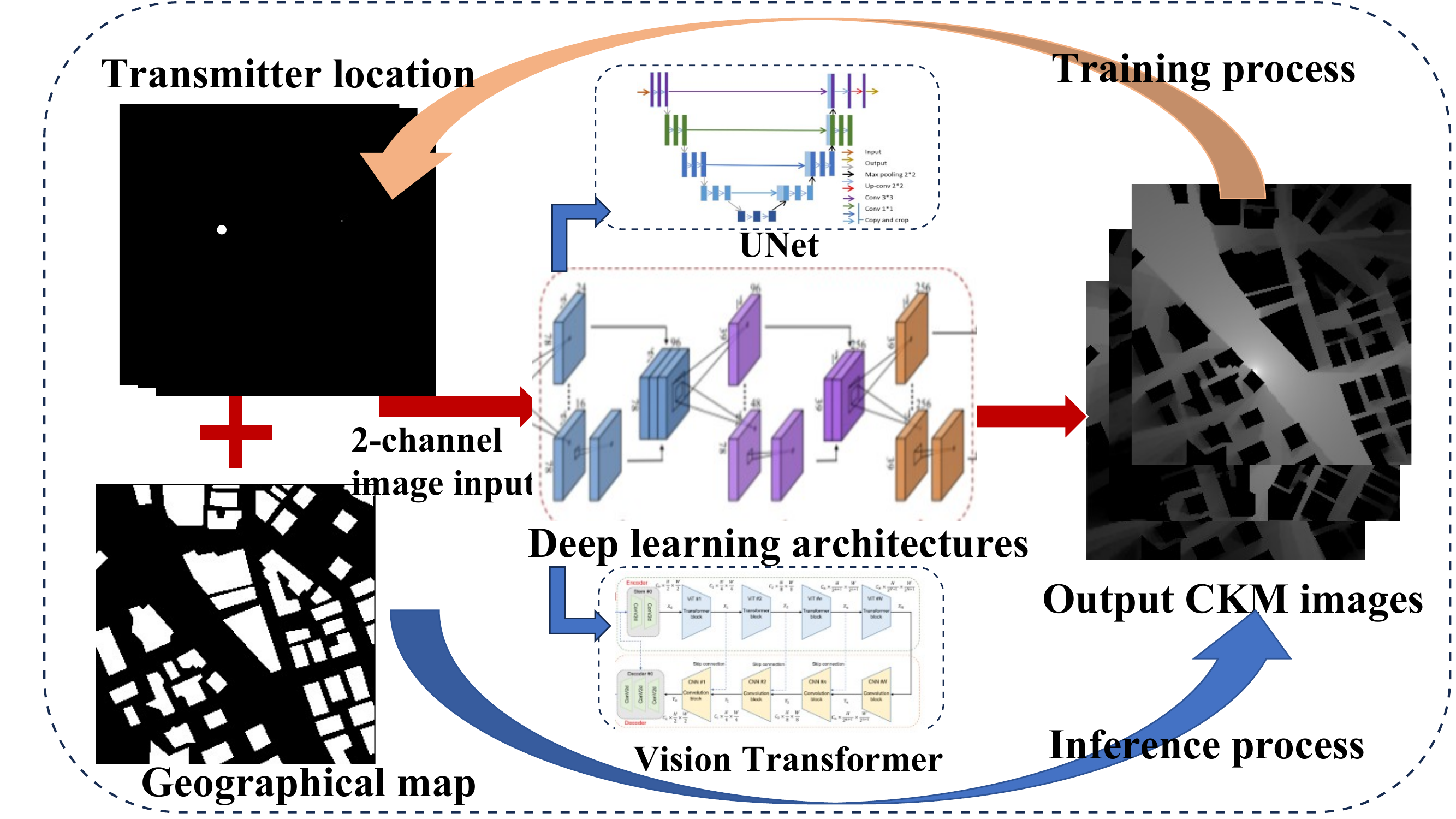}\caption{\label{fig:CNN--and-Transformer-based}Deep learning-based architecture
for CKM construction.}
\end{figure}

Conventional deep learning-based approaches, such as RadioUNet and
RadioTransformer \cite{levie2021radiounet,li2025radiotransformer},
use relationships between image features and channel knowledge to
construct CKMs across different scenarios. The deep learning-based
CKM construction framework is illustrated in Fig. \ref{fig:CNN--and-Transformer-based},
in which the CKM is represented as a 2D image. This image-based representation
naturally enables the use of established computer vision techniques
for feature extraction, denoising, super-resolution, and generative
reconstruction of fine-grained CKMs.

As shown in Fig.\, \ref{fig:CNN--and-Transformer-based}, the CKM
construction framework takes two kinds of environmental inputs: geographical
maps that capture large-scale features such as roads and buildings,
and Tx location masks that specify the spatial distribution of base
stations within the area of interest. These inputs are fed into a
dual-branch deep learning architecture, with each branch dedicated
to processing one modality and extracting its respective spatial and
contextual features. During the joint training phase, both branches
are trained in a supervised manner using high-fidelity CKMs obtained
from ray tracing simulations. At the inference stage, feature-level
fusion combines the localized precision from the Tx branch with the
broad contextual awareness from the geometry map branch, enabling
reliable CKM construction in previously unseen environments. Two representative
implementations are RadioUNet \cite{levie2021radiounet} and RMTransformer
\cite{li2025radiotransformer}, which adopt different architectural
paradigms: a CNN-based U-Net for effective multi-scale spatial feature
extraction, and a vision Transformer (ViT) for capturing long-range
dependencies and global contextual information, respectively. 

Specifically, RadioUNet adopts a UNet-based encoder–decoder architecture
with skip connections to preserve fine-grained spatial details during
reconstruction \cite{levie2021radiounet}. The model is trained on
the large-scale RadioMapSeer dataset, comprising 56, 000 urban propagation
maps by using the mean squared error (MSE) as the training loss. Input
data consist of static urban geometry maps and Tx location masks,
optionally augmented with vehicle maps to capture dynamic shadowing
effects. The framework achieves inference speeds suitable for real-time
deployment on standard central processing units (CPUs). By incorporating
physics-informed priors, RadioUNet effectively learns spatial pathloss
patterns with low reconstruction error in gray-level representation.
Furthermore, simulation-to-measurement transfer learning enhances
generalization capability, yielding noticeably lower reconstruction
errors compared with conventional baselines.

On the other hand, RMTransformer employs a ViT backbone tailored for
CKM construction \cite{li2025radiotransformer}. The model divides
input maps into non-overlapping patches, encodes each patch into a
latent embedding, and applies multi-head self-attention to capture
long-range spatial dependencies. Training is carried out on the RadioMapSeer
dataset with an MSE loss, using input modalities such as urban geometry
maps, Tx location masks, and optionally dynamic vehicle maps. This
design enables RMTransformer to model global contextual relationships
that are often overlooked by convolution-based architectures. The
inference process is sufficiently fast on CPUs to support near real-time
applications. The model attains low gray-level reconstruction errors
and demonstrates superior performance over RadioUNet in challenging
scenarios, such as environments with irregular building layouts or
sparse Tx deployments. Transfer learning from simulated to measured
data further improves generalization, reducing reconstruction errors
compared with RadioUNet. 

While CNN- and Transformer-based CKM models achieve notable reconstruction
accuracy, they still face important challenges. Their dependence on
discretized CKM maps introduces quantization errors, limiting the
recovery of fine-scale propagation features. In addition, these approaches
normally operate in 2D, overlooking vertical propagation and complex
3D structural effects. In addition, they require large, diverse training
datasets, and performance often drops severly in unseen environments. 

\subsection{Generative AI for Multi-task CKM Construction}

In parallel, generative AI models apply data-driven priors to improve
construction quality from sparse or noisy measurement, without relying
on environment maps or computationally expensive ray-tracing simulations
\cite{wu2025ckmimagenet}. Generative AI provides a powerful solution
for CKM construction, directly learning channel distributions from
data to address the ill-posed recovery based on sparse and noisy measurement.
This data-driven approach reduces dependence on costly environmental
priors and supports multiple tasks, such as denoising, inpainting,
and super-resolution, within a unified framework. Its ability to implicitly
model complex channel–environment relationships reduces the need for
extensive measurements, enhancing the practicality and scalability
of CKM construction for intelligent 6G networks.

\begin{figure}[tbh]
\centering

\includegraphics[scale=0.2]{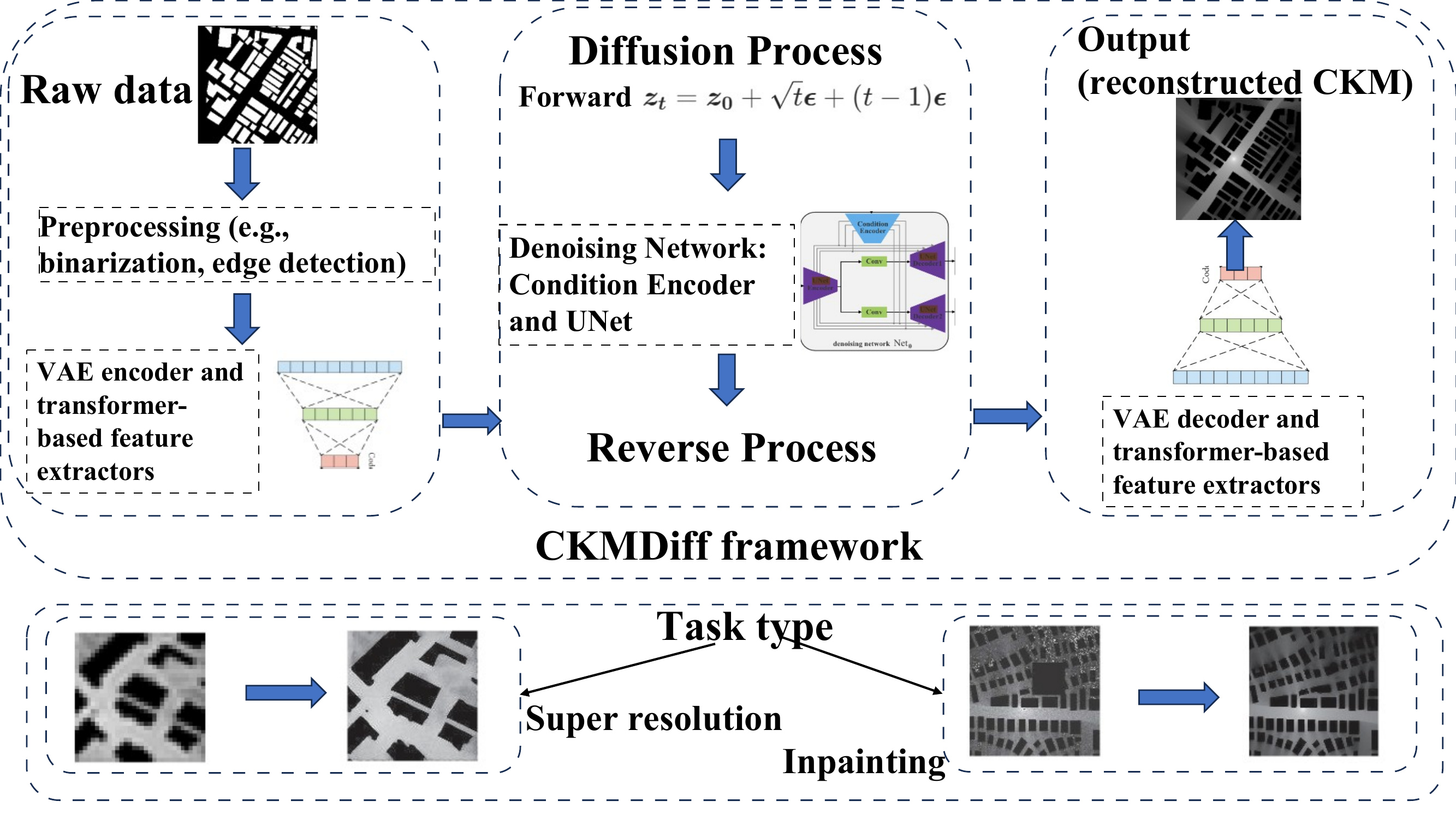}\caption{\label{fig:CKMDiff-framework-and}CKMDiff framework and task type.}

\end{figure}

Based on this idea, CKMDiff is proposed as a conditional decoupled
diffusion model (DDM) integrating a variational autoencoder (VAE)
and Swin Transformer components \cite{fu2025ckmdiff}. As shown in
Fig. \ref{fig:CKMDiff-framework-and}, the VAE encoder maps CKM from
pixel space to latent space, reducing diffusion complexity. The diffusion
process is split into image attenuation and noise enhancement stages.
In the reverse process, a modified UNet with dual decoders predicts
both attenuation and noise terms, conditioned on observed channel
data. A Swin-B encoder extracts multi-level features from observations,
fused with image features at each UNet decoder level to guide reconstruction.
Task-specific preprocessing is applied: zero filling for inpainting,
and direct low-resolution inputs for super-resolution.

CKMDiff is trained on the CKMImageNet dataset containing 30,000 channel
gain images at 128\texttimes 128 resolution \cite{fu2025ckmdiff}.
Training uses a batch size of 48 with 400,000 iterations for general
tasks, and a reduced batch size of 16 with 300,000 iterations for
super-resolution, where UNet channels are increased to process richer
high-resolution features. Experiments demonstrate that CKMDiff consistently
outperforms CNN- and model-based baselines. In super-resolution, it
achieves up to 31.66\% lower root mean squared error (RMSE) than the
RadioUNet, highlighting its ability to capture intrinsic CKM structures
and generalize from limited data. These results confirm its robustness
and adaptability in diverse wireless environments.

Generative AI enables flexible CKM construction by learning implicit
channel–environment relationships, yet faces notable limitations.
Most generative AI is confined to 2D B2X scenarios, missing vertical
propagation effects in complex 3D environments. These schemes also
demand large, diverse datasets, making data acquisition costly. Furthermore,
the multi-step diffusion sampling introduces latency that limits real-time
applicability. Addressing these challenges calls for 3D-aware CKM
architectures, data-efficient learning, and faster inference to fully
realize its potential in 6G systems.

\section{WRF for CKM Construction}

Motivated by the success of radiance field rendering, particularly
NeRF and 3DGS, in computer graphics, the use of WRF methods for CKM
construction has garnered growing research interest. In the optical
domain, radiance field rendering aims to synthesize novel views of
a scene from sparse observations by learning its continuous 3D radiance
representation. NeRF maps 3D coordinates and viewing directions to
color and density for photorealistic reconstructions, while 3DGS uses
anisotropic Gaussian primitives to approximate the radiance field,
enabling much faster rendering with comparable fidelity.

In the wireless domain, WRF representation integrates electromagnetic
physics with computational neural networks to characterize RF signal
propagation, including reflection, diffraction, and scattering in
complex multipath environments, for which NeRF and 3DGS offer key
approaches. Neural radio-frequency radiance fields (NeRF\texttwosuperior )
\cite{zhao2023nerf2} extends optical NeRF to the radio frequency
(RF) domain by modeling complex-valued signals with MLPs, while WRF-GS
\cite{wen2024wrf} uses 3D Gaussian primitives to capture signal amplitude,
phase, and attenuation, enabling efficient scene representation. Both
frameworks employ geometric projection (e.g., Mercator) to project
3D field representations onto 2D receiver planes, followed by splatting-based
spatial spectrum synthesis. 

\subsection{$\textrm{NeRF}^{2}$}

\begin{figure}[tbh]
\centering

\includegraphics[scale=0.2]{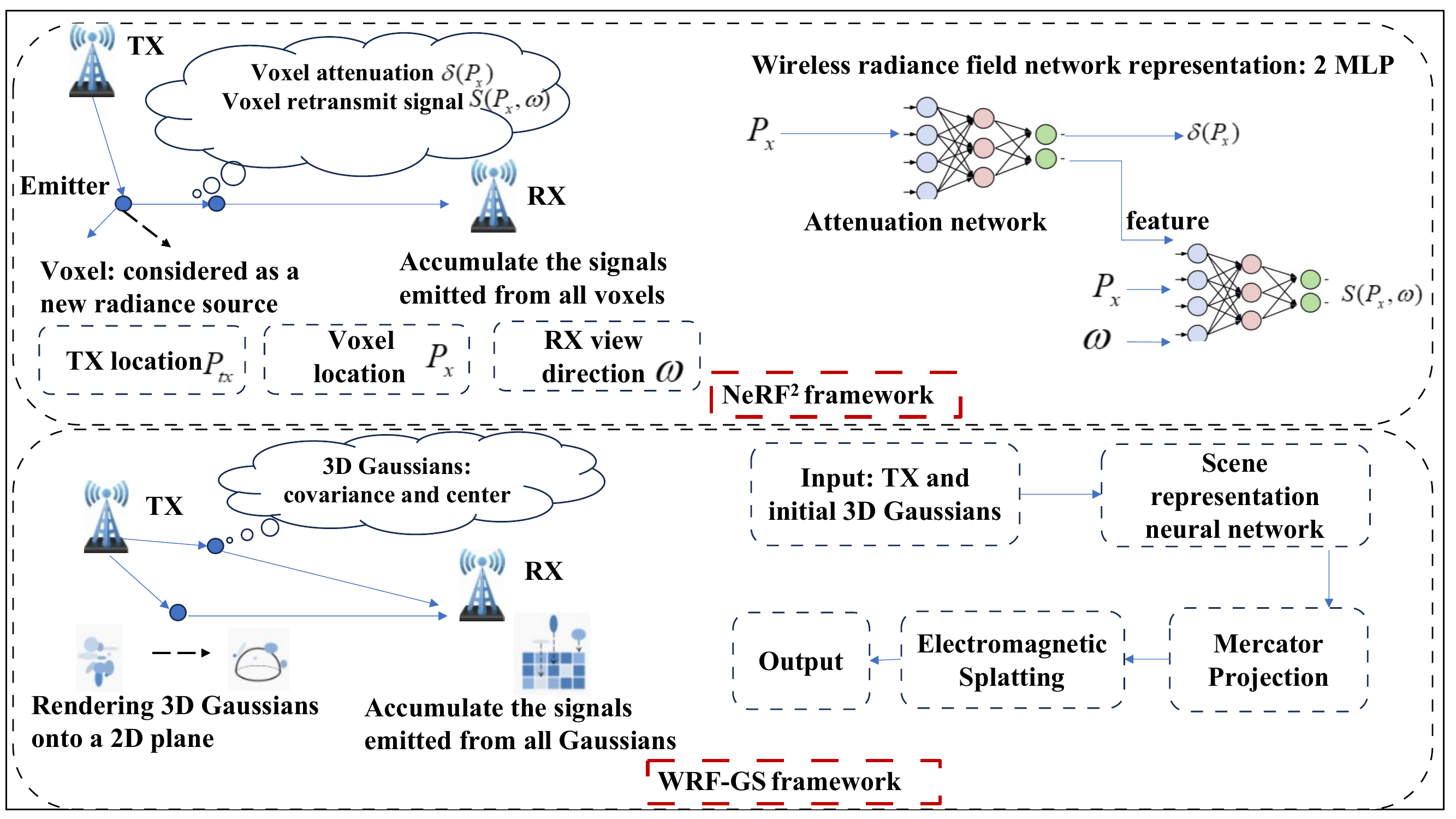}\caption{\label{fig:WRF-based-representation}NeRF\texttwosuperior{} and WRF-GS
frameworks.}
\end{figure}

NeRF\texttwosuperior{} \cite{zhao2023nerf2} extends the optical neural
radiance field framework to the RF domain by representing wireless
propagation as a continuous volumetric scene function. Leveraging
sparse RF measurement, NeRF\texttwosuperior{} can infer the received
complex-valued signals, accounting for both amplitude and phase at
arbitrary locations without explicit 3D scene reconstruction. As shown
in Fig. \ref{fig:WRF-based-representation}, NeRF\texttwosuperior{}
employs a hybrid modeling strategy combining physical principles (e.g.,
Friis equation) with learned neural representations. The core computation
consists of attenuation and radiance MLPs that map Tx positions, voxel
attributes, and view directions to complex-valued signals, explicitly
modeling both amplitude and phase, instead of omitting phase information
in optical NeRF. This design allows NeRF\texttwosuperior{} to implicitly
capture reflections, diffractions, and other multipath phenomena without
requiring explicit 3D environment reconstructions, thereby overcoming
a central bottleneck of conventional CKM construction. During inference,
the trained neural networks predict voxel-specific attenuation and
direction-dependent re-emission parameters from the Tx location and
viewing direction. These outputs are then aggregated through an electromagnetic
procedure to synthesize spatial spectra across Rx planes. By embedding
the physical model into the rendering pipeline, NeRF\texttwosuperior{}
achieves high extrapolation accuracy with minimal measurement, effectively
removing the dependency on complete geometry scans that limit conventional
CKM construction.

The experiment in \cite{zhao2023nerf2} across 14 diverse scenes show
that NeRF\texttwosuperior{} substantially outperforms existing methods,
such as fingerprinting-based interpolation and VAE. For localization
tasks, it significantly reduces the median error and standard deviation
compared with fingerprinting-based interpolation approaches. For channel
prediction, it achieves notable performance improvements over state-of-the-art
VAE-based models, enabling more accurate beamforming. These gains
stem from NeRF\texttwosuperior ’s integration of physical propagation
patterns into data-driven learning, whereas traditional CKM methods
often rely purely on statistical fitting.

Nevertheless, NeRF\texttwosuperior{} training remains computationally
expensive, restricting its real-time applicability. Its static-scene
assumption also degrades performance in dynamic environments with
moving obstacles. Moreover, like many data-driven CKM approaches,
cross-scene generalization is limited, requiring retraining for each
deployment. 

\subsection{WRF-GS}

WRF\nobreakdash-GS is another type of wireless channel modeling framework,
which reconstructs the WRF explicitly using 3DGS rather than voxel\nobreakdash-based
neural radiance fields. By representing scattering points in the environment
as virtual Txs with Gaussian primitives, WRF\nobreakdash-GS can directly
embed attenuation and signal strength attributes, project them onto
the perception plane of an antenna array, and synthesize spatial spectra
through a fast electromagnetic splatting process. Compared with NeRF\texttwosuperior ’s
volumetric rendering, WRF\nobreakdash-GS offers markedly lower computational
complexity and faster rendering, making it better suited for latency\nobreakdash-sensitive
scenarios. Its sample efficiency is also higher, as the compact Gaussian
representation captures propagation characteristics with fewer training
measurement \cite{wen2024wrf}.

As shown in Fig. \ref{fig:WRF-based-representation}, the key innovation
of WRF-GS is its adaptation of the 3DGS technique for WRF reconstruction—a
step that effectively bridges optical scene representation and RF
channel modeling. In this framework, 3D Gaussian primitives are treated
as virtual Txs. A scenario representation network captures signal
strength and attenuation, then projects these primitives onto the
receiver’s antenna plane via a Mercator projection to synthesize spatial
spectra. This design directly supports CKM’s core objective, i.e.,
explicitly modeling the propagation paths and their interactions with
physical obstacles.

Experimental evaluations in \cite{wen2024wrf} demonstrate that WRF\nobreakdash-GS
consistently outperforms a variety of representative baselines in
both spatial spectrum synthesis and practical prediction tasks, such
as VAE and NeRF\texttwosuperior . Across these diverse categories,
WRF\nobreakdash-GS integrates physics\nobreakdash-aware scene modeling
with compact 3D Gaussian primitives to deliver state\nobreakdash-of\nobreakdash-the\nobreakdash-art
accuracy, millisecond\nobreakdash-level latency, and high sample
efficiency, enabling fast, scalable, and data\nobreakdash-efficient
reconstruction of wireless radiation fields for real\nobreakdash-time
CKM construction.

WRF\nobreakdash-GS adopts a one\nobreakdash-way modeling framework
that assumes a fixed Rx location. This formulation does not account
for dynamic switching of transceiver locations, making it more suited
to scenarios with predetermined Tx-Rx configurations and less aligned
with X2X CKM applications that require flexible bidirectional modeling
across arbitrary Tx\nobreakdash-Rx pairs.

\subsection{Bidirectional Wireless Gaussian Splatting (BiWGS)}

\begin{figure}[tbh]
\centering\includegraphics[scale=0.35]{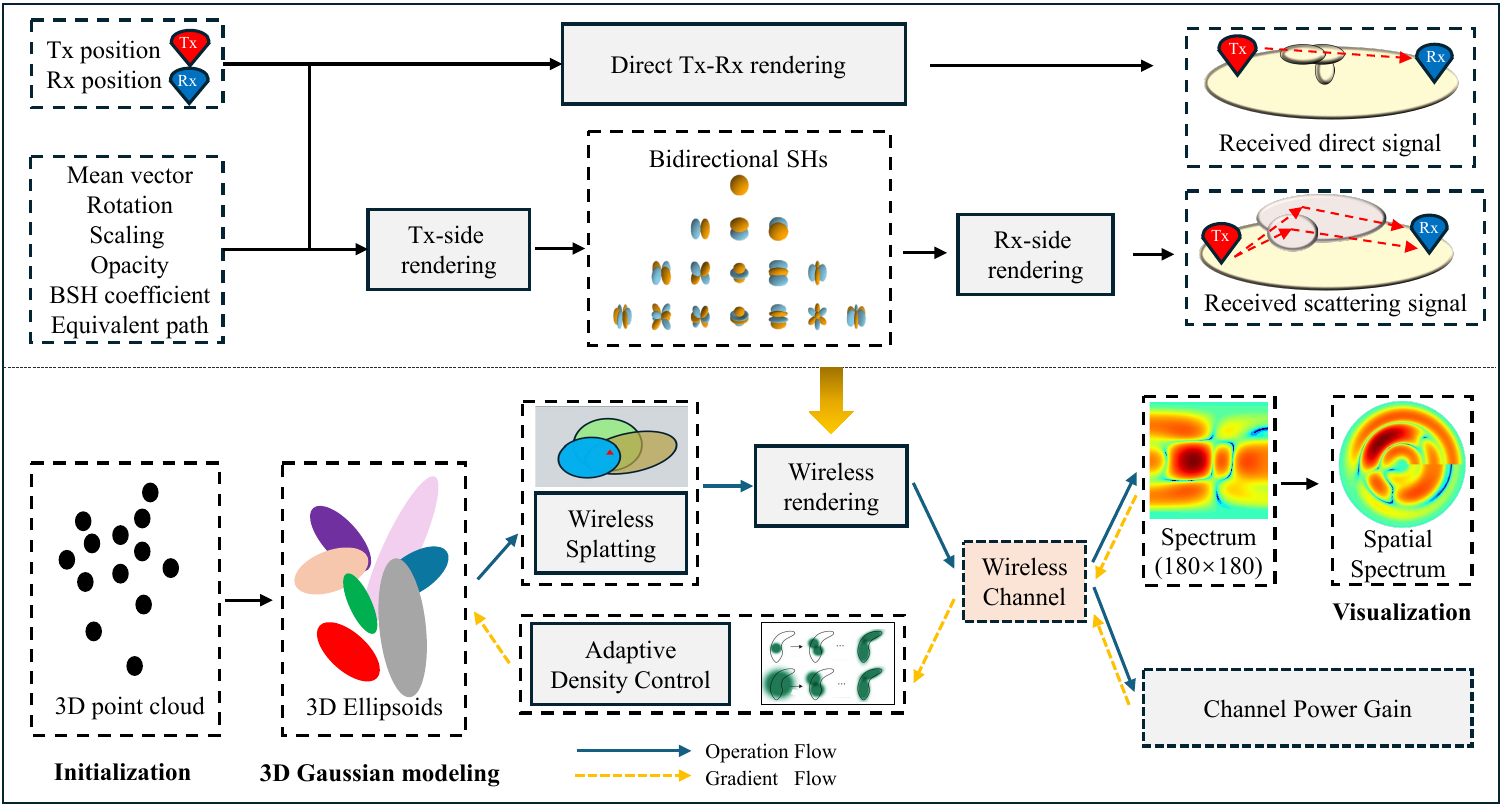}\caption{\label{fig:B2X-CKM-construction}BiWGS framework \cite{zhou}.}

\end{figure}

Recently, a 6D X2X CKM construction method has been proposed in \cite{zhou}.
To extend WRF\nobreakdash-based modeling from conventional B2X scenarios
to fully general X2X CKM construction, a novel framework called BiWGS
is developed, which is capable of modeling channel knowledge for arbitrary
Tx–Rx position pairs in 3D space. BiWGS augments the classical 3DGS
paradigm with a physically grounded bidirectional scattering mechanism,
in which each Gaussian ellipsoid represents a virtual scatterer cluster
characterized by both geometric attributes and electromagnetic response.
Through this representation, BiWGS captures key bidirection wireless
channel phenomena, including multipath propagation, distance\nobreakdash-dependent
amplitude attenuation, phase rotation, and channel reciprocity constraints.
By embedding these physics\nobreakdash-aware parameters into the
splatting and rendering stages, BiWGS transcends the unidirectional
nature of previous WRF\nobreakdash-based approaches, thereby more
faithfully replicating the bidirectional signal interactions inherent
in real environments. This design effectively bridges the gap between
optical radiance\nobreakdash-field modeling and wireless channel
representation, enabling high\nobreakdash-fidelity 6D X2X CKM construction
across diverse environments while retaining the computational efficiency
required for practical deployment.

As shown in Fig. \ref{fig:B2X-CKM-construction}, the wireless rendering
pipeline separates Tx\nobreakdash- and Rx\nobreakdash-side attenuation
calculations, deriving per\nobreakdash-path amplitude loss and phase
distortion from ellipsoid opacity and equivalent path length, while
modeling bidirectional scattering via a factorized complex coefficient
fitted with real and imaginary bidirectional scattering harmonic (BSH)
terms and enforced reciprocity through spherical harmonic parity constraints
for physically consistent symmetry.

The splatting stage adapts the computer\nobreakdash-graphics projection
process to the RF domain by mapping 3D ellipsoids onto virtual reception
planes defined for specific AoA samples, replacing perspective projection
with parallel projection to preserve distance\nobreakdash-invariant
scaling. This allows Gaussian primitives to act as physics\nobreakdash-aware
basis functions, in which opacity models amplitude attenuation, coordinates
encode distance\nobreakdash-dependent loss, and phase terms capture
multipath interference. Training is implemented via using a joint
loss function combining dB\nobreakdash-scale spatial spectrum MSE
and channel power\nobreakdash-gain, ensuring balanced learning of
angular scattering patterns and overall link budget. An adaptive Gaussian\nobreakdash-density
control strategy (including cloning, splitting, and pruning) continually
refines ellipsoid placement in high\nobreakdash-gradient regions,
accelerating convergence and improving representation fidelity. BiWGS
overcomes the unidirectional limitations of prior WRF\nobreakdash-based
schemes and enables high\nobreakdash-fidelity, computationally efficient
construction of 6D X2X CKMs.

\begin{figure}[tbh]
\centering\includegraphics[scale=0.5]{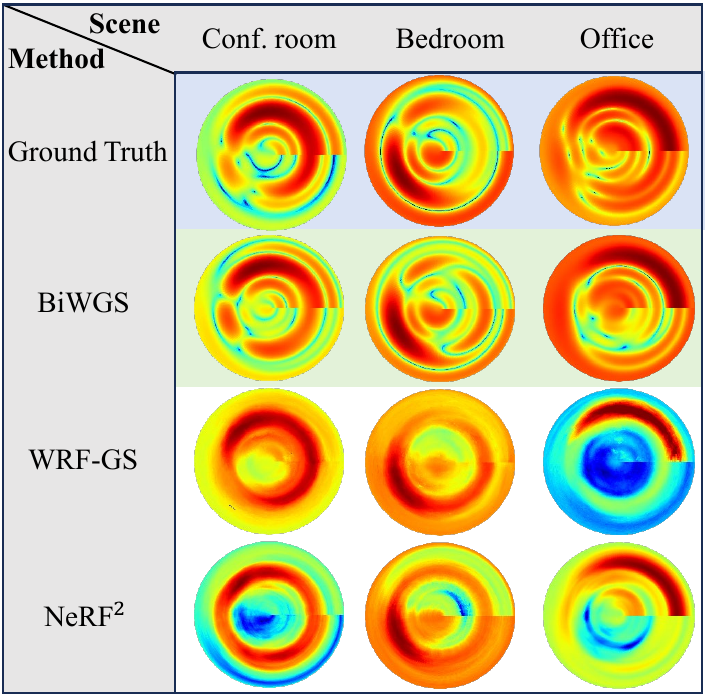}\caption{\label{fig:Comparative-visualization-of}Comparative visualization
of spatial spectrum predictions for 3D CKM construction \cite{zhou}.}

\end{figure}

The performance of BiWGS has been evaluated in \cite{zhou} in a controlled
3D wireless environment generated by the NVIDIA Sionna ray\nobreakdash-tracing
simulator, adopting both 3D B2X CKM and 6D X2X CKM construction tasks.
For 3D B2X CKM settings, BiWGS achieves structural similarity index
measure (SSIM) scores on par with the state-of-the-art WRF-GS ($0.6787$
vs. $0.6858$) and surpasses it in learned perceptual image patch
similarity (LPIPS) by a large margin ($0.4565$ vs. $0.5715$), producing
sharper and less ambiguous spectrum reconstructions. Fig. \ref{fig:Comparative-visualization-of}
shows the visualization of spatial spectrum predictions for 3D CKM
under three representative environments. Compared with the ground
truth, WRF-GS and NeRF\texttwosuperior{} generate spectra with blurred
lobes and spatial ambiguities, leading to the loss of fine-grained
multipath structures. In contrast, BiWGS produces results that are
visually much closer to the ground truth, with sharper spectral patterns
and well-preserved secondary components. This observation is consistent
with the quantitative improvements, especially in the LPIPS metric,
and further confirms the capability of BiWGS to provide physically
meaningful and perceptually faithful reconstructions of 3D CKMs. In
terms of efficiency, BiWGS’s parallel\nobreakdash-projection splatting
and BSH fitting incur longer training than WRF\nobreakdash-GS yet
remain far faster than NeRF\texttwosuperior ’s voluminous ray\nobreakdash-marching.
At inference, BiWGS synthesizes a spatial spectrum in \textasciitilde 0.01\,s—slightly
slower than WRF\nobreakdash-GS but orders\nobreakdash-of\nobreakdash-magnitude
faster than NeRF\texttwosuperior , making it practical for interactive
CKM queries. 

Furthermore, 6D X2X CKM construction example is also provided in \cite{zhou}
with varying Tx–Rx positions. BiWGS significantly outperforms a classical
MLP baseline in channel power gain prediction, reducing mean absolute
errors by up to $54\%$ in challenging non-line-of-sight conditions
and preserving high prediction accuracy for unseen Tx positions. These
results demonstrate that BiWGS achieves high-fidelity, generalizable,
and computationally efficient CKM construction, marking a substantial
advancement in environment-aware wireless communication modeling.

\section{Future Directions}

Despite the rapid progress in CKM construction, several critical challenges
remain for future exploration.

\textbf{Real-time updates and computational latenc}y: Although existing
CKM construction approaches capture fine-grained spatial variations
with high fidelity, their reliance on dense sampling and rendering
can introduce significant computational latency. In dynamic wireless
environments where channel states may vary on millisecond timescales,
such delays hinder real-time CKM updates. This motivates the development
of lightweight neural architectures, adaptive resolution schemes,
and incremental update mechanisms that can refresh the CKM without
recomputing the entire model (e.g., radiance field), thereby meeting
the stringent latency requirements of 6G applications.

\textbf{Cross-domain and cross-frequency generalization}: CKMs often
show degraded performance when applied across different frequency
bands, propagation conditions, or deployment scenarios. This limitation
arises from the weak coupling between purely data-driven representations
and underlying physical propagation laws. A promising direction is
to embed electromagnetic propagation equations into the model architecture,
enabling physics-informed neural representations that naturally generalize
across domains and frequency ranges, thus improving generalization
in complex network environments.

\textbf{Multi-modal fusion for CKM construction}: Purely radio-based
CKM estimation may suffer from blind spots in areas with sparse measurement
or severe multipath. Integrating complementary sensing modalities,
such as LiDAR for precise geometry, radar for material characterization,
or visual cameras for semantic context, could enhance reconstruction
accuracy and resilience. Multi-modal fusion frameworks that operate
at both feature and decision levels can exploit the strengths of each
modality, yielding CKM representations that are richer, more complete,
and less sensitive to single-modality degradation.

\textbf{Scalability and hardware efficiency}: Deploying existing CKM
construction models in large-scale networks with a large number of
BSs and Tx imposes stringent scalability requirements. Current high-fidelity
models often demand considerable memory, compute resources, and specialized
hardware, which may not be practical at scale. Research into large-scale
scene representations, distributed processing strategies, and hardware-aware
model compression is crucial to achieving architectures that scale
gracefully while maintaining accuracy in resource-constrained environments.

\textbf{Standardized datasets and evaluation benchmarks}: The absence
of standardized CKM datasets and evaluation protocols makes it difficult
to compare different approaches objectively and hinders reproducibility.
A shared benchmark that includes diverse propagation scenarios, frequency
bands, and measurement densities would accelerate research and help
bridge the gap between algorithm development and real-world deployment.
Such datasets should be accompanied by well-defined metrics for accuracy,
latency, and scalability, ensuring fair comparison and guiding future
innovations in the field.

\section{Conclusion }

This article presented a comprehensive review of CKM construction
technologies, tracing the evolution from interpolation-based schemes
to deep learning–enhanced image processing and WRF frameworks. Interpolation
methods offer computational simplicity but rely on stationarity and
dense sampling, limiting performance in complex, high-dimensional
environments. Image-based and generative AI approaches enhance accuracy
and multi-task flexibility, yet facing challenges in real-time inference,
3D propagation modeling, and cross-scene generalization. WRF frameworks
mark a paradigm shift by representing wireless propagation as a spatially
continuous radiance field, enabling high-resolution 3D B2X and 6D
X2X 6D CKM construction. Across these developments, key trends include
the integration of real-time updates, multi-modal fusion, lightweight
architectures, and scalable deployment strategies. 

\bibliographystyle{IEEEtran}
\bibliography{IEEEabrv,IEEEexample,Magzine/myref}

\begin{thebibliography}{10}
\providecommand{\url}[1]{#1}
\csname url@samestyle\endcsname
\providecommand{\newblock}{\relax}
\providecommand{\bibinfo}[2]{#2}
\providecommand{\BIBentrySTDinterwordspacing}{\spaceskip=0pt\relax}
\providecommand{\BIBentryALTinterwordstretchfactor}{4}
\providecommand{\BIBentryALTinterwordspacing}{\spaceskip=\fontdimen2\font plus
\BIBentryALTinterwordstretchfactor\fontdimen3\font minus
  \fontdimen4\font\relax}
\providecommand{\BIBforeignlanguage}[2]{{%
\expandafter\ifx\csname l@#1\endcsname\relax
\typeout{** WARNING: IEEEtran.bst: No hyphenation pattern has been}%
\typeout{** loaded for the language `#1'. Using the pattern for}%
\typeout{** the default language instead.}%
\else
\language=\csname l@#1\endcsname
\fi
#2}}
\providecommand{\BIBdecl}{\relax}
\BIBdecl

\bibitem{zeng2021toward}
Y.~Zeng and X.~Xu, ``Toward environment-aware {6G} communications via channel
  knowledge map,'' \emph{IEEE Wireless Commun.}, vol.~28, no.~3, pp. 84--91,
  Mar. 2021.

\bibitem{howmuchdata}
X.~Xu and Y.~Zeng, ``How much data is needed for channel knowledge map
  construction?'' \emph{{IEEE} Trans. Wireless Commun.}, vol.~23, no.~10, pp.
  13\,011--13\,021, May 2024.

\bibitem{zeng2024tutorial}
Y.~Zeng, J.~Chen, J.~Xu, D.~Wu, X.~Xu, S.~Jin, X.~Gao, D.~Gesbert, S.~Cui, and
  R.~Zhang, ``A tutorial on environment-aware communications via channel
  knowledge map for {6G},'' \emph{IEEE Commun. Surv. Tutor.}, vol.~26, no.~3,
  pp. 1478--1519, Feb. 2024.

\bibitem{wu2023environment}
D.~Wu, Y.~Zeng, S.~Jin, and R.~Zhang, ``Environment-aware hybrid beamforming by
  leveraging channel knowledge map,'' \emph{{IEEE} Trans. Wireless Commun.},
  vol.~23, no.~5, pp. 4990--5005, Oct. 2023.

\bibitem{10962296}
H.~Sun, L.~Zhu, and R.~Zhang, ``Channel gain map estimation for wireless
  networks based on scatterer model,'' \emph{{IEEE} Trans. Wireless Commun.},
  pp. 1--1, early Access, 2025.

\bibitem{sun2022propagation}
H.~Sun and J.~Chen, ``Propagation map reconstruction via interpolation assisted
  matrix completion,'' \emph{{IEEE} Trans. Signal Process.}, vol.~70, pp.
  6154--6169, Dec. 2022.

\bibitem{li2022channel}
K.~Li, P.~Li, Y.~Zeng, and J.~Xu, ``Channel knowledge map for environment-aware
  communications: {EM} algorithm for map construction,'' in \emph{2022 IEEE
  Wireless Commun. Net. Conf. (WCNC)}.\hskip 1em plus 0.5em minus 0.4em\relax
  IEEE, 2022, pp. 1659--1664.

\bibitem{I2IInpainting}
Z.~Jin, L.~You, J.~Wang, X.-G. Xia, and X.~Gao, ``{An I2I} inpainting approach
  for efficient channel knowledge map construction,'' \emph{{IEEE} Trans.
  Wireless Commun.}, vol.~24, no.~2, pp. 1415--1429, Dec. 2025.

\bibitem{wu2025ckmimagenet}
Z.~Wu, D.~Wu, S.~Fu, Y.~Qiu, and Y.~Zeng, ``{CKMImageNet: A} dataset for
  {AI}-based channel knowledge map towards environment-aware communication and
  sensing,'' \emph{arXiv preprint arXiv:2504.09849}, 2025.

\bibitem{fu2025ckmdiff}
S.~Fu, Y.~Zeng, Z.~Wu, D.~Wu, S.~Jin, C.-X. Wang, and X.~Gao, ``{CKMDiff}: {A}
  generative diffusion model for ckm construction via inverse problems with
  learned priors,'' \emph{arXiv preprint arXiv:2504.17323}, 2025.

\bibitem{zhao2023nerf2}
X.~Zhao, Z.~An, Q.~Pan, and L.~Yang, ``{Nerf2: Neural} radio-frequency radiance
  fields,'' in \emph{Proc. 29th Annu. Int. Conf. Mobile Comput. Netw.}, 2023,
  pp. 1--15.

\bibitem{wen2024wrf}
C.~Wen, J.~Tong, Y.~Hu, Z.~Lin, and J.~Zhang, ``{WRF-GS}: {Wireless} radiation
  field reconstruction with {3D} gaussian splatting,'' \emph{IEEE INFOCOM
  2025-IEEE Conference on Computer Communications}, pp. 1--10, 2025.

\bibitem{zhou}
J.~Zhou, C.~Hu, G.~Wu, Z.~Ren, H.~Hu, J.~Zhang, R.~Zhang, and J.~Xu, ``{6D}
  channel map construction via bidirectional {Gaussian} splatting,''
  \emph{arXiv preprint arXiv:2510.26166}, 2025.

\bibitem{levie2021radiounet}
R.~Levie, {\c{C}}.~Yapar, G.~Kutyniok, and G.~Caire, ``{RadioUNet: Fast} radio
  map estimation with convolutional neural networks,'' \emph{{IEEE} Trans.
  Wireless Commun.}, vol.~20, no.~6, pp. 4001--4015, Feb. 2021.

\bibitem{li2025radiotransformer}
Y.~Li, C.~Zhang, W.~Wang, and Y.~Huang, ``{RadioTransformer: Accurate} radio
  map construction and coverage prediction,'' \emph{arXiv preprint
  arXiv:2501.05190}, 2025.

\end{thebibliography}

\end{document}